\documentclass{PoS}

\usepackage{enumerate}
\usepackage{graphicx}

\PoS{PoS(LAT2005)208}

\title{Study of 1/m corrections in HQET }

\ShortTitle{Study of 1/m corrections in HQET }

\author{\speaker{Shunsuke Negishi}
\\
        Department of Physics, Kyoto University, Kyoto 606-8501, Japan\\
        E-mail: \email{negishi@phys.h.kyoto-u.ac.jp}}

\author{Hideo Matsufuru\\
        High Energy Accelerator Research Organization (KEK), Tsukuba 305-0801, Japan\\
        E-mail: \email{hideo.matsufuru@kek.jp}}

\author{Tetsuya Onogi\\
        Yukawa Institute for Theoretical Physics, Kyoto University, Kyoto 606-8502, Japan\\
        E-mail: \email{onogi@yukawa.kyoto-u.ac.jp}}

\author{Takashi Umeda\\
        Brookhaven National Laboratory, Upton, New York, 11973, USA\\
        E-mail: \email{tumeda@quark.phy.bnl.gov}}

\abstract{We report our exploratory study on 
the matching condition of HQET with QCD including $1/m$ 
corrections. We introduce a new observable from the dependence of the heavy-light 
effective energy on the twisted boundary condition parameter $\theta$, which 
could be used to match the kinetic term $\vec{D}^2/2m$. 
Carrying out quenched QCD simulations for fixed lattice spacing 
in small volumes with O($a$)-improved Wilson fermions, we study 
the $1/m$ dependence of this observable, from which 
the static limit and $1/m$ coefficient can be extracted. 
We also compare our preliminary result with HQET.}

\FullConference{XXIIIrd International Symposium on Lattice Field Theory\\
25-30 July 2005\\
Trinity College, Dublin, Ireland}

\begin{document}

\section{Introduction}
One of the primary goals in particle physics is to determine 
fundamental parameters of the standard model such as the quark masses 
and the Cabibbo-Kobayashi-Maskawa matrix elements 
(e.g. $V_{\rm td},V_{\rm ts},V_{\rm tb}$)  in order to test the standard model 
and  find a clue to the physics beyond the standard model.
However for this determination in addition to the experimental 
inputs from CLEO-c \cite{1}, BaBar \cite{2} and BELLE \cite{3}, 
we need non-perturbative evaluation of hadronic observables from 
the first principles of QCD. Above all, simulation of lattice QCD 
is one of the most effective approaches.

In lattice QCD, the typical lattice cutoff accessible in the 
present computers is $a^{-1}\approx 1\sim 3{\rm GeV}$. 
But in the case of heavy quarks such as the bottom quark 
whose mass is $m\approx 4{\rm GeV}$ the quark mass in lattice unit 
is so large $ma\approx 1\sim 4$ that the conventional 
lattice fermion action has no control over the discretization errors.
In order to avoid this problem which has been raised of the heavy-quark calculation, 
we must take the following approaches: \vspace{-6pt}
\begin{enumerate}[(a)]
\item conventional fermions with much finer lattice spacing, \label{a}\vspace{-7pt}
\item non-relativistic QCD and their variations, \label{b}\vspace{-7pt}
\item anisotropic lattices \cite{5,6},\label{c}\vspace{-7pt}
\item heavy-quark effective theory(HQET) \cite{7,8,9}.\label{d}\vspace{-4pt}
\end{enumerate}
However (\ref{a}) is still not practical in high precision computation due to computational cost and
(\ref{b}) has theoretical uncertainty owing to the perturbative
matching, while (\ref{c}) and (\ref{d}) have potential of providing 
formalism for high precision computation.
In this report, we will give our exploratory study on (\ref{d}) HQET.

HQET is a theory that reproduces the low energy mode of a heavy quark in
the static limit \cite{10,11}, where the action is a systematic 
expansion in inverse powers of the mass
\begin{eqnarray}
S_{\rm HQET}=a^4\sum_x\left\{{\cal L}_{\rm stat}(x)+\sum_{\nu=1}^n{\cal L}^{(\nu)}(x)\right\},\hspace{1.5cm}\label{1-1}\\
{\cal L}_{\rm stat}(x)=\bar\psi_{\rm h}\left(\nabla_0^*+\delta m\right)\psi_{\rm h}(x),\label{1-3}\ 
{\cal L}^{(\nu)}(x)=\sum_i\omega_i^{(\nu)}{\cal L}_i^{(\nu)}(x),\\
\delta m={\cal O}\left(\frac{1}{m}\right),\ 
\omega_1^{(1)}=\frac{1}{m}+{\cal O}\left(\frac{1}{m^2}\right),\ 
\omega_2^{(1)}=\frac{1}{m}+{\cal O}\left(\frac{1}{m^2}\right),\label{1-2}\\
{\cal L}_1^{(1)}(x)=\bar{\psi_{\rm h}}\left(-\frac{1}{2}{\bf\sigma\cdot B}\right)\psi_{\rm h},\ 
{\cal L}_2^{(1)}(x)=\bar{\psi_{\rm h}}\left(-\frac{1}{2}{\bf D}^2\right)\psi_{\rm h}.
\end{eqnarray}
If one treats in $1/m$ corrections perturbatively to a fixed order,  
the theory is a renormalizable theory with well-defined continuum limit:
We presume it can accept up until next-to-leading order terms of $1/m$ and can have $1\sim2\%$ accuracy.
However HQET is an effective theory with unknown coefficients 
for $1/m$ corrections terms and the ultraviolet behavior is essentially 
different from QCD.  Therefore, one has to match HQET with QCD 
by matching conditions defined by a set of physical observables
\begin{eqnarray}
\left<{\cal O}_{\rm QCD}(1/z)\right>=\left<{\cal O}_{\rm HQET}(1/z)\right>,\ z\equiv ML\gg1\ 
({\cal O};{\rm an\ observable}),\label{1-4}
\end{eqnarray}
where $M$ is a renormalization group invariant mass \cite{12}.
This can practically be done in a small volume $L\approx0.2{\rm fm}$ where we take the small lattice spacing within reasonable numerical cost:\linebreak[4]
$m_{\rm b}a\ll1$. 
Then by computing the step-scaling function \cite{11}, 
we can evolve the lattice HQET into coarser lattice where physical observables can be computed in large volume.
In fact these two steps have been carried out for $\delta m_{\rm q}$
and the axial current $A^{\rm QCD}_\mu=A^{\rm HQET}_\mu$ in the static
limit, from which the following results are obtained
\begin{eqnarray}
F^{\rm stat}_{\rm B_s}=253\pm45{\rm MeV}\ \cite{14},\ 
\bar{m}_{\rm b}=4.12\pm0.07\pm0.04{\rm MeV}\ \cite{11}. 
\end{eqnarray}
For a high precision calculation, non-perturbative calculation 
of the $1/m$ correction is required. For this purpose it is 
necessary to match the coefficients of the $1/m$ corrections 
in the action or operators to begin with. 

The matching of HQET with QCD is composed of three steps:\vspace{-6pt}
\begin{enumerate}[(a)]
\setcounter{enumi}{4}
\item to execute the matching condition (\ref{1-4}),\label{e}\vspace{-7pt}
\item to match coefficients of $1/m$-correction terms from Eq. (\ref{1-4}),\label{f}\vspace{-7pt}
\item to evaluate the step scaling functions.\label{g}\vspace{-4pt}
\end{enumerate}
In this report we give study on (\ref{e}) as the first step for (\ref{f}) and also computing the static limit; 
especially on the search for new observables which are efficient for the determination of $1/m$ term. 

\section{Observables}
In Ref.~\cite{13}, Heitger \textit{et al.} propose using two-point correlation functions in small finite volume as the matching condition (\ref{1-4}).
Thus we can evaluate some HQET parameters by combining observables in varied kinematical conditions 
and execute heavy quark lattice simulations of both QCD and HQET in $L_0\approx 0.2{\rm fm}$.

Two-point correlation functions are defined by
\begin{eqnarray}
f_{\rm A}(x_0) = -\frac{1}{2}\int d^3{\bf y}d^3{\bf z}
\left< A_0\bar\zeta_{\rm b}({\bf y})\gamma_5\zeta_{\rm l}({\bf z}) \right>,\ 
k_{\rm V}(x_0) = -\frac{1}{6}\sum_k\int d^3{\bf y}d^3{\bf z}
\left< V_k\bar\zeta_{\rm b}({\bf y})\gamma_k\zeta_{\rm l}({\bf z})
\right>, 
\end{eqnarray}
where $A_\mu(x)=\bar\psi_{\rm l}(x)\gamma_\mu\gamma_5\psi_{\rm b}(x)$,
$V_\mu(x)=\bar\psi_{\rm l}(x)\gamma_\mu\psi_{\rm b}(x)$, and
$\bar\zeta$, $\zeta$ are the boundary fields at $x_0=0$.
Defining effective energies as
\begin{eqnarray}
E^{\rm eff}_{\rm PS}(x_0,\theta) = -\frac{d}{dx_0}\ln[f_{\rm A}(x_0)],\ 
E^{\rm eff}_{\rm V}(x_0,\theta) = -\frac{d}{dx_0}\ln[k_{\rm V}(x_0)],
\end{eqnarray}
we can determine various observables in kinematical conditions such as
\begin{eqnarray}
\Gamma_{\rm av}=\frac{1}{4}\left(E^{\rm eff}_{\rm PS}\left(\frac{T}{2}\right)+3E^{\rm eff}_{\rm V}\left(\frac{T}{2}\right)\right)\label{2-6},\
\Delta\Gamma=E^{\rm eff}_{\rm V}\left(\frac{T}{2}\right)-E^{\rm eff}_{\rm PS}\left(\frac{T}{2}\right),\\
\Xi=\frac{1}{4}\left(E^{\rm eff}_{\rm PS}\left(\frac{T}{4}\right)+3E^{\rm eff}_{\rm V}\left(\frac{T}{4}\right)\right)
-\frac{1}{4}\left(E^{\rm eff}_{\rm PS}\left(\frac{T}{2}\right)+3E^{\rm eff}_{\rm V}\left(\frac{T}{2}\right)\right)
\end{eqnarray}
to assess $\delta m,\ \omega_1^{(1)},\ {\rm and\ }\omega_2^{(1)}$ respectively, defined in Eq. (\ref{1-2}).

It is known that the kinetic term of $1/m$ correction $\omega_2^{(1)}$ is difficult to evaluate it with good accuracy,
because of the cancellation between power divergences of effective energies at $T/2$ and $T/4$, 
which causes large systematic errors.
Therefore it is fruitful to look for alternative observables for consistency checks and pursuit of more efficiency.
In this work we propose an alternative observable:
\begin{eqnarray}
\Xi^{\rm new}=\Gamma_{\rm av}(1/z)|_{\theta=1.0}-\Gamma_{\rm av}(1/z)|_{\theta=0.5},\label{2-9}
\end{eqnarray}
where $\theta$ is defined by twisted boundary conditions
\begin{eqnarray}
\psi(x+\hat kL)=e^{i\theta}\psi(x),\ \bar\psi(x+\hat
kL)=\bar\psi(x)e^{-i\theta} 
&&(k=1,2,3,\ T=L,\ \theta=0.5,1.0).
\end{eqnarray}

\section{Simulation methods}
Simulations are carried out both for QCD and HQET on a quenched $12^4$
lattice with the standard plaquette gauge action at $\beta=6/g^2_0=7.4802$.
We take the Schr\"odinger functional boundaries \cite{16,17} with $C=C'=0$.
The SF-boundary coefficients are 
$c_{\rm t}^{\rm2\textrm{-}loop}=1-0.089g_0^2-0.030g_0^4$ \cite{15}.
256 configurations are accumulated by the pseudo-heat-bath algorithm each 
separated by 200 Monte Carlo sweeps. 
Errors are estimated by the standard jackknife method.

In QCD, we use $O(a)$-improved Wilson fermion with the nonperturbative 
clover coefficient $c_{\rm SW}$ ~\cite{18} and the SF boundary 
conditions $P_+\psi(x)|_{x_0=0}=\rho$,  $P_-\psi(x)|_{x_0=T}=0$ \cite{22}.
Six values of hopping parameters are taken for the heavy quark 
which correspond to $z\equiv ML =$ 3.0, 3.8, 5.15, 6.0, 6.6, 9.0  
whereas the hopping parameter for the light quark is set to the 
critical value $(\kappa_{\rm c}=0.133961)$ following Ref.~ \cite{21}.

In HQET, we used the static-limit action ${\cal L}_{\rm stat}(x)$
(\ref{1-3}) with the boundaries for the heavy quark. 
In order to improve the numerical precision, 
we adopted gauge fields in $\nabla^*_0$ of ${\cal L}_{\rm stat}(x)$
are smeared by the HYP links \cite{20} with
$(\alpha_1,\alpha_2,\alpha_3)=$ (1.0, 1.0, 0.5) ~\cite{23,24}. 
The light quark is the same as in QCD. 

\section{Results}
\begin{figure}[ht]
\begin{minipage}{.45\linewidth}
\hspace{-1.25cm}
\rotatebox{270}{
\includegraphics[width=5.75cm]{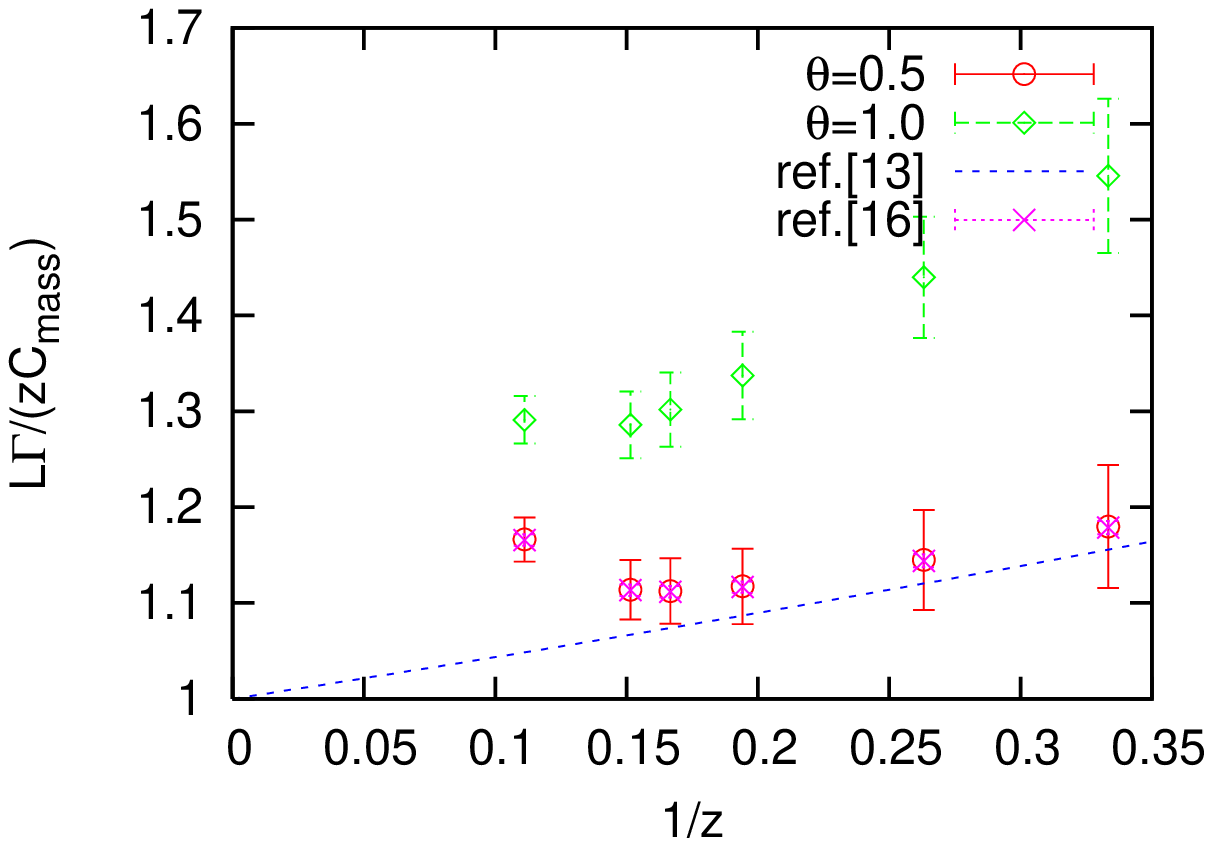}
}
\caption{The averages of the effective energy at $\theta=0.5$ (circle
symbol) and $\theta=\ 1.0$ (diamond). The cross symbols 
and the dashed line denote the 
data for $\theta=0.5$ at $\beta=7.4802$ ~\cite{21} 
and in the continuum limit ~\cite{13} respectively.}
\label{fig1}
\end{minipage}
\begin{minipage}{1.65cm}
\ 
\end{minipage}
\begin{minipage}{.45\linewidth}
\hspace{-1.4cm}
\rotatebox{270}{
\includegraphics[width=5.75cm]{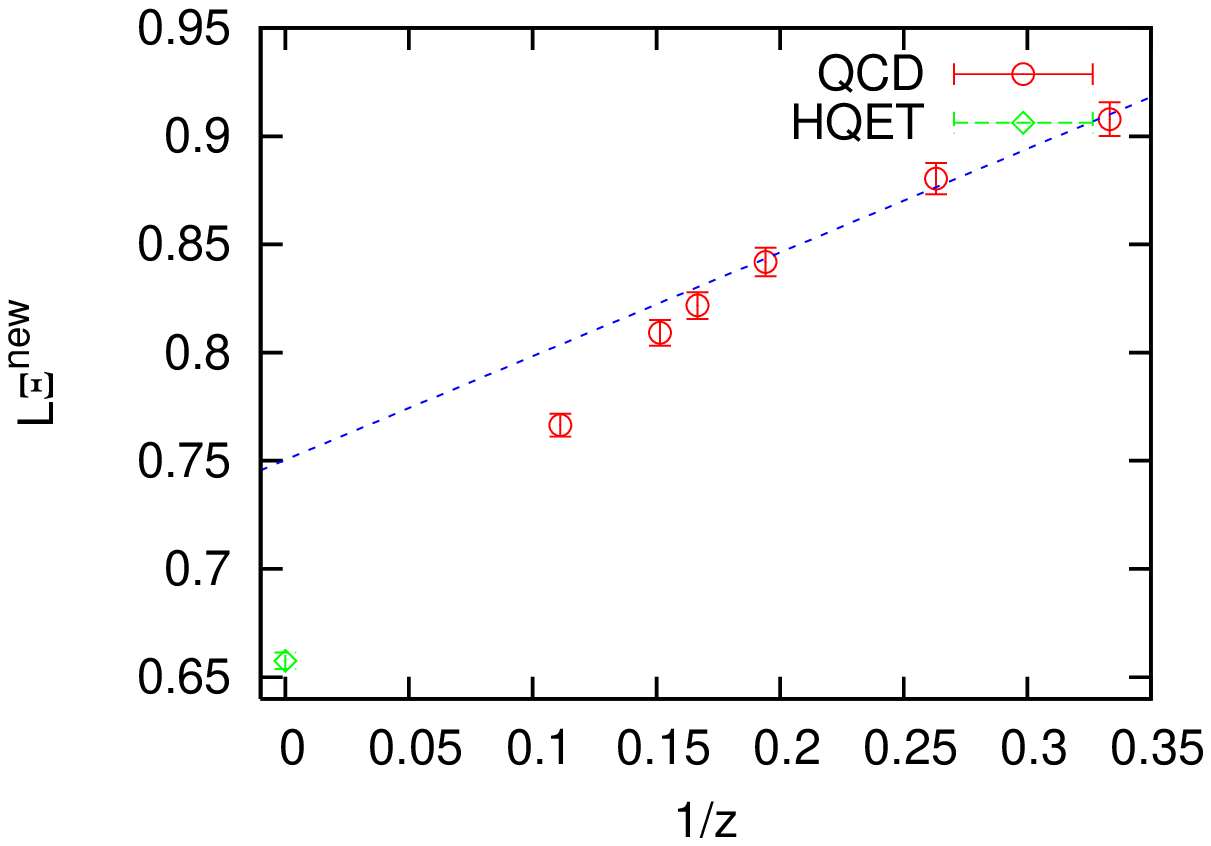}
}
\caption{$1/z$ dependence of the new observable $L\Xi^{\rm new}$.
The circle symbols denote our data for QCD and the diamond denotes the HQET 
result. The dashed line is the linear fit using three lighter points.}
\label{fig2}
\end{minipage}
\end{figure}

Fig.~\ref{fig1} shows the average effective energies 
$\Gamma_{\rm av}$ (\ref{2-6}) at $\theta=0.5,\ 1.0$ with matching 
condition
\begin{eqnarray}
&L\Gamma_{\rm av}\stackrel{M\to\infty}{\sim}C_{\rm
mass}(M/\Lambda_{\rm\overline{MS}})z+{\cal O}((1/z)^0)\
\cite{13},\label{3-1}
\end{eqnarray}
where $C_{\rm mass}$ is the matching function between QCD and
HQET~\cite{13}  and $M/\Lambda_{\rm\overline{MS}}$ is given in Ref.~\cite{12}.
We reproduced consistent result with $\theta=0.5$ 
at $\beta=7.4802$ in Ref.~\cite{21}. As is obvious from 
Fig.~\ref{fig1}, the data for $\theta=0.5$ at 
$\beta=7.4802$ deviate only by $2\sim3\%$ for 
$z=3.0,\ 3.8,\ 5.15 $, whereas for $z\equiv ML =9.0$ 
the deviation is of 10\% order due to the discretization error. 
Although there is no scaling study, we find 
the $1/z$ dependence of the result for $\theta=1.0$ also shows 
qualitatively similar behavior.  In fact, our data at  
$z=3.0,\ 3.8,\ 5.15$ seem to approach the static limit value 
predicted as in Eq.~(\ref{3-1}).
On the other hand, the data for $z=9.0$ seem to deviate from the 
$1/m$ scaling significantly. There is need of a 
stringent test in further studies, since this deviation is probably due to discretization 
errors.

Fig.~\ref{fig2} shows the $1/z$ dependence of the observable $\Xi^{\rm new}$. 
Owing to the reparameterization invariance~\cite{19}, 
the renormalization factor is not needed for this observable.  
Assuming that the data for $z=3.0,\ 3.8,\ 5.15$ are very close to the
continuum limit, we made a linear fit in $1/z$ as
$L\Xi^{\rm new} = a_0 +a_1/z$. Our preliminary results are 
$a_0 = 0.750(12)$ and $a_1 =0.479(47)$, where the errors are statistical
only.  The data for $z=9.0$ again deviate from the
$1/z$ scaling, 
which is probably due to the discretization error. 
The extrapolation of QCD results does approach to the HQET 
result towards the static limit,  
although it gives higher values than the HQET result 
at $\beta=7.4802$. Further study is needed to see if 
this discrepancy vanishes in the continuum limit. 
Nevertheless, it is encouraging that this observable has 
clear $1/z$ dependence so that it has sensitivity 
to determine the coefficient of $1/m$-correction terms 
$\omega_2^{(1)}$ in HQET. 
 
\section{Conclusions}
We have proposed the new observable $\Xi^{\rm new}$ defined as 
the difference of the effective energy at $\theta=0.5,\ 1.0$ 
for extracting the coefficients of the kinetic term in HQET. 
Our simulation at $\beta=7.4802$ suggests that it has 
qualitatively correct HQET scaling and also good sensitivity 
to $1/m$-correction term.

{\bf Acknowledgements: }
The simulation has been done at the Yukawa Institute Computer Facility and
on a supercomputer (NEC SX-5) at Research Center for Nuclear Physics, Osaka University.

\end{document}